\documentclass{IEEEtran4PSCC}

\ifCLASSINFOpdf

\else

\fi

\usepackage[cmex10]{amsmath}
\usepackage{graphics,graphicx,epic,eepic,epsfig,times,units} % epstopdf

\usepackage{siunitx}
\DeclareSIUnit \VAr {VAr} %volt-ampere reactive  
\DeclareSIUnit \VA {VA} %volt-ampere 
\DeclareSIUnit \rad {Radians} % Radians

\usepackage{cite}

\usepackage{cleveref}
\crefformat{equation}{(#2#1#3)}
\crefrangeformat{equation}{(#3#1#4)--(#5#2#6)}
\crefmultiformat{equation}{(#2#1#3)}%
{,~(#2#1#3)}{, (#2#1#3)}{,~(#2#1#3)}

\usepackage[ruled,vlined]{algorithm2e}

\usepackage{xcolor}

\usepackage{algpseudocode}
\usepackage{algcompatible}
\algrenewcommand\alglinenumber[1]{\scriptsize #1:}

\usepackage{tabularx}

\newcommand{\reals}{{\mbox{\bf R}}}

\DeclareMathOperator*{\argmin}{arg\,min}

\usepackage{psfrag}

\newtheorem{definition}{Definition}
\let\olddefinition\definition
\renewcommand{\definition}{\olddefinition\normalfont}

\allowdisplaybreaks

\begin{document}
	
	\title{Accelerated Methods for the SOCP-relaxed Component-based Distributed Optimal Power Flow}

	\author{\IEEEauthorblockN{Sleiman~Mhanna,~\emph{MIEEE},
			Gregor~Verbi\v{c},~\emph{Senior~MIEEE,}
			and~Archie~C.~Chapman,~\emph{MIEEE}}}
	
	\maketitle
	
	\begin{abstract}
		In light of the increased focus on distributed methods, this paper proposes two accelerated subgradient methods and an adaptive penalty parameter scheme to speed-up the convergence of ADMM on the component-based dual decomposition of the second-order cone programming (SOCP) relaxation of the OPF. This work is the first to apply an adaptive penalty parameter method along with an accelerated subgradient method together in one scheme for distributed OPF. This accelerated scheme is demonstrated to reach substantial speed-ups, as high as 87\%, on real-world test systems with more than 9000 buses, as well as on other difficult test cases.
	\end{abstract}
	

	\section*{Notation}
	
	\addcontentsline{toc}{section}{Notation}
	\subsection{Input data and operators}
	%\small
	\begin{IEEEdescription}[\IEEEsetlabelwidth{$p_{gi}/q_{gi}$}\IEEEusemathlabelsep]
		\item[$\mathcal{B}$] Set of buses in the power network.
		\item[$\mathcal{B}_{i}$] Set of buses connected to bus $i$.
		\item[$b^\text{sh}_{i}$] Shunt susceptance (p.u.) at bus $i$.
		\item[$g^\text{sh}_{i}$] Shunt conductance (p.u.) at bus $i$.
		\item[$b^\text{ch}_{ij}$] Charging susceptance (p.u.) in the $\pi$-model of line $ij$.
		\item[$c0_{gi}$] Constant coefficient ($\SI{}{\$}$) term of generator $g$'s cost function.
		\item[$c1_{gi}$] Coefficient ($\SI{}{\$\per\mega\watt}$) of the linear term of generator $g$'s cost function.
		\item[$c2_{gi}$] Coefficient ($\SI{}{\$\per\mega\watt\squared}$) of the quadratic term of generator $g$'s cost function.
		\item[$\mathcal{G}$] Set of all generators $(g,i)$ in the power network such that $g$ is the generator and $i$ is the bus connected to it.
		\item[$\mathcal{G}_{i}$] Set of all generators connected to bus $i$.
		\item[$\mathrm{j}$] Imaginary unit.
		\item[$\mathcal{L}$] Set of all branches $ij$ where $i$ is the ``from'' bus.
		\item[$\mathcal{L}_{t}$] Set of all branches $ji$ where $j$ is the ``to'' bus.
		\item[$p_{i}^{\text{d}}/q_{i}^{\text{d}}$] Active/reactive power demand ($\SI{}{\mega\watt}/\SI{}{\mega\VAr}$) at bus $i$.
		\item[$\overline{s}_{ij}$] Apparent power rating ($\SI{}{\mega\VA}$) of line $ij$.
		\item[$\underline{\theta}_{ij}^{\Delta}$] Lower limit of the difference of voltage angles of buses $i$ and $j$.
		\item[$\overline{\theta}_{ij}^{\Delta}$] Upper limit of the difference of voltage angles of buses $i$ and $j$.
		\item[$\theta_{i}^{\text{shift}}$] Phase shift ($\SI{}{\rad}$) of phase shifting transformer connected between buses $i$ and $j$ ($\theta_{i}^{\text{shift}}=0$ for a transmission line).
		\item[$\tau_{ij}$] Tap ratio magnitude of phase shifting transformer connected between buses $i$ and $j$ ($\tau_{ij}=1$ for a transmission line).
		\item[$T_{ij}$] Complex tap ratio of a phase shifting transformer ($T_{ij}=\tau_{ij}\mathrm{e}^{\mathrm{j} \theta_{i}^{\text{shift}}}$).
		\item[$Y_{ij}$] Series admittance (p.u.) in the $\pi$-model of line $ij$.
		\item[$\Im\left\{\bullet\right\}$] Imaginary value operator.
		\item[$\Re\left\{\bullet\right\}$] Real value operator.
		\item[$\underline{\bullet}/\overline{\bullet}$] Minimum/maximum magnitude operator.
		\item[$\left|\bullet\right|$] Magnitude operator/Cardinality of a set.
		\item[$\bullet^*$] Conjugate operator.
		\item[$k$] Iteration number.
		\item[$\rho$] ADMM penalty parameter.
	\end{IEEEdescription}
	
	\subsection{Decision variables}
	\begin{IEEEdescription}[\IEEEsetlabelwidth{$p_{gi}/q_{gi}$}\IEEEusemathlabelsep]
		\item[$p_{gi}/q_{gi}$] Active/reactive power ($\SI{}{\mega\watt}/\SI{}{\mega\VAr}$) generation of generator $g$ at bus $i$.
		\item[$p_{gi_{(i)}}$] Duplicate of $p_{gi}$ at bus $i$.
		\item[$q_{gi_{(i)}}$] Duplicate of $q_{gi}$ at bus $i$.
		\item[$p_{ij}/q_{ij}$] Active/reactive power ($\SI{}{\mega\watt}/\SI{}{\mega\VAr}$) flow along branch $ij$.
		\item[$p_{ij_{(i)}}$] Duplicate of $p_{ij}$ at bus $i$.
		\item[$q_{ij_{(i)}}$] Duplicate of $q_{ij}$ at bus $i$.
		\item[$V_{i}$] Complex phasor voltage (p.u.) at bus $i$ ($V_{i}=\left| V_{i} \right| \angle \theta_{i}=v_{i} \angle \theta_{i}$).	
		\item[$\boldsymbol{\lambda}$] Vector of Lagrange multipliers.
	\end{IEEEdescription}
	
%	\subsection{Acronyms}
%	\begin{IEEEdescription}[\IEEEsetlabelwidth{$p_{gi}/q_{gi}$}\IEEEusemathlabelsep]
%		\item[AC] Alternating current.
%		\item[ADMM] Alternating direction method of multipliers.
%		\item[GNLP] Global nonlinear programming.
%		\item[KKT] \emph{Karush-Kuhn-Tucker} conditions.
%		\item[NLP] Nonlinear programming.
%		\item[OCD] Optimality conditions decomposition.
%		\item[OPF] Optimal power flow.
%	\end{IEEEdescription}

	\section{Introduction}\label{sec:Introduction}
	
	Up to this day, most optimization and control algorithms in power systems, such as the optimal power flow (OPF), are computed in a centralized fashion. With the increasing penetration of distributed energy resources however, the feasibility of the centralized computation paradigm is at stake for four main reasons. First, collecting all the required information from these DERs to centrally compile an OPF problem instance entails substantial communication overhead. Second, this information is private to the owners of these DERs and accessing it would raise privacy concerns. Third, most of these DERs require the use of mixed-integer variables to model them accurately. Their presence results in a mixed-integer nonlinear program (MINLP), which further increases the computational complexity of the AC OPF problem. Therefore, the resulting large-scale MINLP may be intractable if solved centrally to optimality. Fourth, centralized schemes are vulnerable to single point of failure or attack. To this end, an increased attention is given to distributed methods as they can be scalable, privacy preserving and robust to single point of failure.
	
	There is a plethora of existing works on distributed OPF. These can be broadly classified into three categories, dual decomposition methods, optimality conditions decomposition (OCD) methods and sparse semidefinite programming (SDP) decomposition methods (see \cite{Molzahn2017_Surveyofdistributedmethods} and \cite{Mhanna2018_Componentbaseddecomposition} for a review). The dual decomposition techniques underlying the dual-decomposition-based distributed OPF methods in the literature can in turn be classified into two categories: region-based decompositions and component-based decompositions. The focus of this study revolves around the latter decomposition techniques because they can distribute the computation across \emph{every} component in the network (generators, transformers, loads, buses, transmission lines etc.) and are flexible enough to incorporate discrete decision variables to suit a wide variety of optimization applications in power systems and future grids. The result of the component-based dual decomposition is a consensus problem that can be solved in a distributed fashion using ADMM. ADMM was first introduced in \cite{Glowinski1975_ADMM} and its convergence was studied in \cite{Gabay1976_ADMMconvergence,Glowinski1989_ALR,Eckstein1992_OntheDRSmethod}.	ADMM is a highly desirable scheme to solve the distributed OPF problem because of its simplicity. However, its practical performance is poor when the problem is ill-conditioned or when high accuracy is required. The convergence of ADMM is also sensitive to the choice of penalty parameter. 
	
	Against this background, and motivated by the electricity industry's real-time decision-making applications, this paper proposes and compares three methods for accelerating the convergence of ADMM on the component-based dual decomposition of the second-order cone programming (SOCP) relaxation of the OPF. In more detail, the first method, called \emph{over-relaxed} ADMM, is a popular variant of ADMM. This method is analyzed in \cite{Eckstein1992_OntheDRSmethod} and \cite{Nishihara_GeneralanalysisofADMM}, and is shown in \cite{Eckstein1994_parallelADMM} and \cite{Ghadimi2015_Optimalparameterselection} to improve the convergence of ADMM. Like the ``vanilla'' ADMM, this method is distributed, as each component communicates with its immediate neighbours, and there is no centralized controller. The second method is an adaptation of the predictor-corrector-type acceleration with restart scheme proposed in \cite{Goldstein2014_FastADMM} (and later in \cite{Kadkhodaie2015_AcceleratedADMM}) for ADMM with weakly convex objectives, which is itself an adaptation of an optimal accelerated gradient descent method initially proposed by Nesterov \cite{Nesterov2004_introductorylectures}. However, in contrast to the over-relaxed ADMM method, this scheme is no longer fully distributed, as the restart rule relies on a combined residual which requires a central controller to compute the global primal and dual residuals. Moreover, since the convergence of ADMM is highly sensitive to the choice of penalty parameters, the third method is an adaptive consensus ADMM in which the penalty parameters are automatically tuned without a central oversight, which is suitable for dynamic network topologies underlying distributed consensus problems. Specifically, adaptive consensus ADMM sets the penalty parameters for each consensus constraint based on the relative magnitudes of the local primal and dual residuals. This method is inspired by the residual balancing scheme, and is demonstrated to achieve substantial speed-ups and relative insensitivity to the initial penalty parameter values and ill-conditioning. More interestingly, the convergence of the first two algorithms is further improved by overlaying them with the adaptive consensus ADMM scheme.
	
	In light of recent promises of recovering feasible solutions from the SOCP relaxation of the OPF \cite{Tian2017_RecoverfeasiblesolutionsSOCP}, the accelerated methods developed in this work are demonstrated on the SOCP-relaxed OPF for various good reasons. First, the SOCP-relaxed OPF is convex, which entails that applying ADMM to solve it comes with convergence guarantees. Another reason for working on the SOCP-relaxed OPF is that this relaxation is shown to be exact in radial networks under some mild conditions \cite{Farivar2013_Branchflowmodel,Gan2015_ExactSOCPradial}. Moreover, in mesh networks, on top of achieving small optimality gaps on many real-world test systems \cite{NESTA},  the SCOP-relaxed OPF can be strengthened even further by bound tightening techniques and tight convex hulls \cite{Chen2015_Boundtightening,Coffrin_Strengtheningwithboundtightening,Kocuk_strongSOCP}. A tightened SOCP-relaxed OPF in mesh
	networks is also used in \cite{Tian2017_RecoverfeasiblesolutionsSOCP} and its solution is used as an initial point for a proposed alternative convex optimization (ACP) OPF algorithm to recover a feasible solution. This ACP algorithm first formulates the OPF problem as a difference-of-convex programming (DCP) problem, then solves the DCP problem by penalty convex-concave procedure (CCP) iteratively. Finally, the component-based dual decomposition applied to the SOCP-relaxed OPF relishes closed-form solutions for the bus and generator subproblems. Problems with closed-form solutions are faster to compute compared to when they are solved using a numerical solver. 
	
	\subsection{Contributions of this work}
	
	In summary, this paper advances the state of the art in the following ways:
	\begin{itemize}
		\item This paper is the first to apply an adaptive penalty parameter method along with an accelerated subgradient method together in one scheme for distributed OPF. 
		\item The methods are implemented on real-world test systems \cite{Josz_ACdataMATPOWER} and other difficult test cases from NESTA v6 \cite{NESTA}, and are demonstrated to reach substantial speed-ups, as high as $88\%$. 
	\end{itemize}
	The methods developed is this paper are readily transferable to other applications in power systems that are based on OPF, such as security constrained unit commitment (SCUC) with contingency constraints and multiple transmission system operators (TSOs), stochastic OPF, probabilistic OPF, and multi-period OPF with demand response (DR), to name a few.
	
%	\subsection{Notation}\label{sec:Notation}
%	
%	All vectors are column vectors unless otherwise specified, and $\boldsymbol{0}$ and $\boldsymbol{1}$ are all-zeros and all-ones vectors (respectively) of length depending on context. The inner product of two vectors $\boldsymbol{x}$, $\boldsymbol{y} \in \reals^n$ is delineated by $\left\langle \boldsymbol{x}, \boldsymbol{y} \right\rangle:=\boldsymbol{x}^{T} \boldsymbol{y}$, where $\boldsymbol{x}^{T}$ is the transpose of $\boldsymbol{x}$. The Euclidean norm of a vector $\boldsymbol{x} \in \reals^n$ is denoted by $\left\|\boldsymbol{x}\right\|:=\sqrt{\left\langle \boldsymbol{x}, \boldsymbol{x} \right\rangle}$ and the nonnegative orthant in $\reals^n$ is denoted by $\reals^n_{+}$.
	
%	\subsection{Organization of the paper}
%	
%	The paper starts with a formal description of the alternative-form OPF in general networks in Section~\ref{sec:OPFproblem}, followed by the component-based dual decomposition in Section~\ref{sec:Componentbaseddecomp}. Section~\ref{sec:Acceleratedmethods} presents two accelerated subgradient methods for ADMM and Section~\ref{sec:AdaptiveADMM} presents an adaptive penalty parameter scheme. Section~\ref{sec:Evaluation} contains the numerical evaluation of the algorithms and Section~\ref{sec:Conclusion} concludes the paper. 
	
	\section{The OPF problem}\label{sec:OPFproblem}
	In a power network, the OPF problem consists of finding the least-cost dispatch of power from generators to satisfy the load at all buses in a way that is governed by physical laws, such as Ohm's Law and Kirchhoff's Law, and other technical restrictions, such as transmission line thermal limit constraints. 
	By letting
	\begin{align}
		w_{i} =\left|V_{i}\right|^2, \ w_{ij}^\mathrm{r}  =\Re\left\{V_{i}V_{j}^*\right\}, \ w_{ij}^\mathrm{i}  =\Im\left\{V_{i}V_{j}^*\right\},
	\end{align}
	the \emph{alternative formulation} of the OPF problem can be written as
	\begin{subequations}\label{eq0:opf}
		\begin{align}
		\underset {\substack{p_{gi},q_{gi},w_{i},w_{ij}^\mathrm{r},w_{ij}^\mathrm{i},\\\theta_{i},p_{ij},q_{ij},p_{ji},q_{ji}}} 
		{\mbox{ minimize}} & \sum_{(g,i)\in \mathcal{G}} f_{gi}\left(p_{gi}\right) & \label{eq0:objective}\\
		\text{ subject to} &  \nonumber \\
		\underline{p}_{gi} & \leq p_{gi} \leq \overline{p}_{gi}, \qquad \qquad \quad   (g,i)\in \mathcal{G} & \label{eq0:Pminmax} \\
		\underline{q}_{gi} & \leq q_{gi} \leq \overline{q}_{gi}, \qquad \qquad \quad   (g,i)\in \mathcal{G} & \label{eq0:Qminmax} \\
		\left|\underline{V}_{i}\right|^2 & \leq w_{i} \leq \left|\overline{V}_{i}\right|^2,  \quad \qquad \qquad  \ \ i \in \mathcal{B} & \label{eq0:Vminmax} \\ 
		\underline{\theta}_{ij}^{\Delta} & \leq \theta_{i}-\theta_{j} \leq \overline{\theta}_{ij}^{\Delta}, \quad \quad  \quad \   (i,j) \in \mathcal{L} & \label{eq0:anglediff} \\
		\sum_{(g,i)\in \mathcal{G}} p_{gi}&-p_{i}^{\text{d}} =  \sum_{j\in \mathcal{B}_{i}} p_{ij} + g^{\text{sh}}_{i}w_{i},  \quad \ \ \ i\in \mathcal{B}  & \label{eq0:KCL1}  \\	
		\sum_{(g,i)\in \mathcal{G}}  q_{gi}&-q_{i}^{\text{d}}=  \sum_{j\in \mathcal{B}_{i}} q_{ij} - b^{\text{sh}}_{i}w_{i},    \quad \ \ \ i\in \mathcal{B}  & \label{eq0:KCL2}  \\
		p_{ij}= g^{\text{c}}_{ij} & \ w_{i}-g_{ij}w_{ij}^\mathrm{r}+ b_{ij}w_{ij}^\mathrm{i}, \ \  \  (i,j) \in \mathcal{L} & \label{eq0:Pij} \\
		q_{ij}= b^{\text{c}}_{ij} & \ w_{i}-b_{ij} w_{ij}^\mathrm{r} - g_{ij} w_{ij}^\mathrm{i}, \ \ \  (i,j) \in \mathcal{L} & \label{eq0:Qij} \\
		p_{ji}= g^{\text{c}}_{ji} & \ w_{j}-g_{ji} w_{ij}^\mathrm{r} - b_{ji} w_{ij}^\mathrm{i}, \ \ \  (i,j) \in \mathcal{L} & \label{eq0:Pji} \\
		q_{ji}=b^{\text{c}}_{ji} & \ w_{j}-b_{ji} w_{ij}^\mathrm{r} + g_{ji} w_{ij}^\mathrm{i}, \ \ \  (i,j) \in \mathcal{L} &\label{eq0:Qji} \\	
		& \hspace{-0.35cm}\left(w_{ij}^\mathrm{r}\right)^2 + \left( w_{ij}^\mathrm{i}\right)^2  = w_{i}w_{j},  \ \  (i,j) \in \mathcal{L}  & \label{eq0:RSOC}  \\ 
		& \hspace{-0.35cm} \theta_{j}-\theta_{i}={\rm atan2}(w_{ij}^\mathrm{i},w_{ij}^\mathrm{r}),  \  (i,j) \in \mathcal{L} &  \label{eq0:atan2} \\
		& \sqrt{p_{ij}^2+q_{ij}^2} \leq \overline{s}_{ij}, \ \ \   (i,j) \in \mathcal{L} \cup \mathcal{L}_{t} & \label{eq0:linethermallimit} 
		\end{align}
	\end{subequations}
	where,  $g^{\text{c}}_{ij}:=\Re\left\{\frac{Y_{ij}^*-\mathrm{j}\frac{b^\text{ch}_{ij}}{2}}{\left|T_{ij}\right|^2}\right\}$, $b^{\text{c}}_{ij}:=\Im\left\{\frac{Y_{ij}^*-\mathrm{j}\frac{b^\text{ch}_{ij}}{2}}{\left|T_{ij}\right|^2}\right\}$, $g_{ij}:=\Re\left\{\frac{Y_{ij}^*}{T_{ij}}\right\}$, $b_{ij}:=\Im\left\{\frac{Y_{ij}^*}{T_{ij}}\right\}$, $g^{\text{c}}_{ji}:=\Re\left\{Y_{ji}^*-\mathrm{j}\frac{b^\text{ch}_{ji}}{2}\right\}$, $b^{\text{c}}_{ji}:=\Im\left\{Y_{ji}^*-\mathrm{j}\frac{b^\text{ch}_{ji}}{2}\right\}$, $g_{ji}:=\Re\left\{\frac{Y_{ji}^*}{T_{ji}^*}\right\}$ and $b_{ji}:=\Im\left\{\frac{Y_{ji}^*}{T_{ji}^*}\right\}$, and $f_{gi}\left(p_{gi}\right):=c2_{gi}\left(p_{gi}\right)^2 +c1_{gi}\left(p_{gi}\right)+c0_{gi}$. The OPF in \eqref{eq0:opf} is a nonconvex nonlinear optimization problem that is proven to be NP-hard \cite{Bienstock2015_NPhardnessofACPF,Lehmann2016_ACfeasibility}. The nonconvexities stem from equality constraint \cref{eq0:RSOC}, which describes the boundary of a rotated second-order cone, and \cref{eq0:atan2} which contains the nonconvex ${\rm atan2}$ function. The SOCP relaxation of the OPF in \eqref{eq0:opf} is obtained by ignoring \cref{eq0:atan2} and relaxing \cref{eq0:RSOC} to  
	\begin{align}
		\left(w_{ij}^\mathrm{r}\right)^2 + \left( w_{ij}^\mathrm{i}\right)^2  \leq w_{i}w_{j}, \  (i,j) \in \mathcal{L}, \label{eq0:RSOCP}
	\end{align}
	which is the convex hull of \eqref{eq0:RSOC}.
	
	\section{Component-based dual decomposition and ADMM}~\label{sec:Componentbaseddecomp} 

\begin{figure}[t]
		\centering{
			\includegraphics[width=90mm] {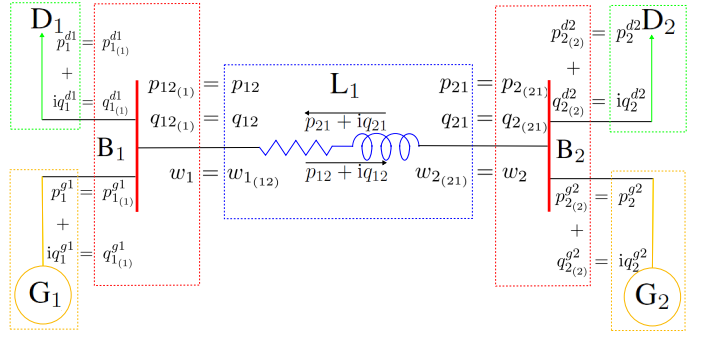}}
		\caption{A 2-bus system showing the duplication of the coupling variables and the resulting component-based decomposition.}
		\label{fig:2Bussystem}
	\end{figure}

	A component-based separability can be bestowed on the SOCP relaxation of \eqref{eq0:opf} by creating copies of the following variables
	\begin{align}
	p_{gi}=& \ p_{gi_{(i)}},  \qquad \qquad \ (g,i)\in \mathcal{G}, \label{eq:conspg} \\
	q_{gi}=& \ q_{gi_{(i)}},  \qquad \qquad \ (g,i)\in \mathcal{G}, \label{eq:consqg} \\
	p_{ij}=& \ p_{ij_{(i)}}, \ \qquad  (i,j) \in \mathcal{L} \cup \mathcal{L}_{t}, \label{eq:consp} \\
	q_{ij}=& \ q_{ij_{(i)}}, \ \qquad  (i,j) \in \mathcal{L} \cup \mathcal{L}_{t}, \label{eq:consq} \\
	w_{i_{(ij)}}=& \ w_{i}, \ \qquad \quad  (i,j) \in \mathcal{L} \cup \mathcal{L}_{t}, \label{eq:consw}
	\end{align}	
	and the SOCP-relaxed OPF problem now becomes
	\begin{subequations}\label{eq1:opf}
		\begin{align}
		\underset {\substack{\boldsymbol{x},\boldsymbol{z},w_{ij}^\mathrm{r},w_{ij}^\mathrm{i}}} 
		{\mbox{ minimize}} & \sum_{(g,i)\in \mathcal{G}} f_{gi}\left(p_{gi}\right) & \label{eq1:objective}\\
		\text{ subject to} &  \text{ \cref{eq0:Pminmax,eq0:Qminmax}, \cref{eq0:linethermallimit}, \cref{eq:conspg,eq:consqg,eq:consp,eq:consq,eq:consw}}  \label{eq1:same} \\
%		\underline{p}_{gi} & \leq p_{gi} \leq \overline{p}_{gi}, \qquad \qquad \qquad \ \ (g,i)\in \mathcal{G} & \label{eq1:Pminmax} \\
%		\underline{q}_{gi} & \leq q_{gi} \leq \overline{q}_{gi}, \qquad \qquad \qquad \ \ (g,i)\in \mathcal{G} & \label{eq1:Qminmax} \\
		\left|\underline{V}_{i}\right|^2 & \leq w_{i_{(ij)}} \leq \left|\overline{V}_{i}\right|^2,  \qquad (i,j) \in \mathcal{L} \cup \mathcal{L}_{t} & \label{eq1:Vminmax} \\ 
		\sum_{(g,i)\in \mathcal{G}} p_{gi_{(i)}}&-p_{i}^{\text{d}} =  \sum_{j\in \mathcal{B}_{i}} p_{ij_{(i)}} + g^{\text{sh}}_{i}w_{i},  \quad \ \ \ i\in \mathcal{B}  & \label{eq1:KCL1}  \\	
		\sum_{(g,i)\in \mathcal{G}}  q_{gi_{(i)}}&-q_{i}^{\text{d}}=  \sum_{j\in \mathcal{B}_{i}} q_{ij_{(i)}} - b^{\text{sh}}_{i}w_{i},    \quad \ \ \ i\in \mathcal{B}  & \label{eq1:KCL2}  \\
		p_{ij}=& \ g^{\text{c}}_{ij} w_{i_{(ij)}}-g_{ij}w_{ij}^\mathrm{r}+ b_{ij}w_{ij}^\mathrm{i},   \  (i,j) \in \mathcal{L} & \label{eq1:Pij} \\
		q_{ij}=& \ b^{\text{c}}_{ij} w_{i_{(ij)}}-b_{ij} w_{ij}^\mathrm{r} - g_{ij} w_{ij}^\mathrm{i},   \  (i,j) \in \mathcal{L} & \label{eq1:Qij} \\
		p_{ji}=& \ g^{\text{c}}_{ji} w_{j_{(ji)}}-g_{ji} w_{ij}^\mathrm{r} - b_{ji} w_{ij}^\mathrm{i},   \  (i,j) \in \mathcal{L} & \label{eq1:Pji} \\
		q_{ji}=& \ b^{\text{c}}_{ji} w_{j_{(ji)}}-b_{ji} w_{ij}^\mathrm{r} + g_{ji} w_{ij}^\mathrm{i},  \  (i,j) \in \mathcal{L} &\label{eq1:Qji} \\	
		& \hspace{-0.5cm}\left(w_{ij}^\mathrm{r}\right)^2 + \left( w_{ij}^\mathrm{i}\right)^2  \leq w_{i_{(ij)}} w_{j_{(ji)}}, \ (i,j) \in \mathcal{L}  & \label{eq1:RSOC} \\
		& \hspace{-1.35cm} {\rm tan}\left( \underline{\theta}_{ij}^{\Delta}\right) w_{ij}^\mathrm{r} \leq w_{ij}^\mathrm{i} \leq {\rm tan}\left(\overline{\theta}_{ij}^{\Delta}\right)w_{ij}^\mathrm{r}, \ (i,j) \in \mathcal{L} & \label{eq1:anglediff} 
		\end{align}
	\end{subequations}
	where 
	\begin{align*}
	\boldsymbol{x}:=&\left[  \left( p_{gi},q_{gi} \right)_{(g,i)\in \mathcal{G}},\left( p_{ij},q_{ij},w_{i_{(ij)}} \right)_{(i,j) \in \mathcal{L} \cup \mathcal{L}_{t}} \right],
	\end{align*}
	and 
	\begin{align*}
	\boldsymbol{z}:=&\left[  \left(p^{g}_{i_{(i)}},q^{g}_{i_{(i)}} \right)_{(g,i)\in \mathcal{G}} , \left(p_{ij_{(i)}},q_{ij_{(i)}}\right)_{(i,j) \in \mathcal{L} \cup \mathcal{L}_{t}},\left( w_{i}\right)_{i \in \mathcal{B}}  \right].
	\end{align*}
	This duplication of the coupling variables along with the resulting component-based decomposition are depicted in Figure~\ref{fig:2Bussystem} for a 2-bus system.
	Let $N_{\boldsymbol{x}}=N_{\boldsymbol{\lambda}}=2 \left| \mathcal{G}\right| + 3 \left| \mathcal{L} \cup \mathcal{L}_{t} \right|$ and $N_{\boldsymbol{z}}=2 \left| \mathcal{G}\right| + 2 \left| \mathcal{L} \cup \mathcal{L}_{t} \right| +\left| \mathcal{B} \right|$. Problem \eqref{eq1:opf} is now of the general form
	\begin{subequations}\label{eq:Generalproblem}
		\begin{align}
		\underset {\substack{\boldsymbol{x} \in \mathcal{X} ,\boldsymbol{z} \in \mathcal{Z}}} 
		{\mbox{ minimize}} & \quad f\left(\boldsymbol{x}\right) + g\left(\boldsymbol{z}\right)  \label{eq:objective}\\
		\text{ subject to} & \quad A\boldsymbol{x}+B\boldsymbol{z}=\boldsymbol{c}, \label{eq:coupling}
		\end{align}
	\end{subequations}
	where $f: \reals^{N_{\boldsymbol{x}}} \rightarrow \reals$ and $g: \reals^{N_{\boldsymbol{z}}} \rightarrow \reals$ are closed convex functions, $A$ is a $N_{\boldsymbol{\lambda}} \times N_{\boldsymbol{x}} $ identity matrix, $B \in \reals^{N_{\boldsymbol{\lambda}} \times N_{\boldsymbol{z}}} $, $\boldsymbol{c} \in \reals^{N_{\boldsymbol{\lambda}}}$,\footnote{Note that $\boldsymbol{c}=\boldsymbol{0}$ in this OPF case.} $\mathcal{X}$ is the feasible set defined by constraints \cref{eq1:same}, \cref{eq1:Vminmax}, \cref{eq1:Pij,eq1:Qij,eq1:Pji,eq1:Qji,eq1:RSOC} and $\mathcal{Z}$ is the feasible set defined by constraints \cref{eq1:KCL1} and \cref{eq1:KCL2}.
	The augmented (partial) Lagrange function of \eqref{eq:Generalproblem} is written as
	\begin{align}\label{eq:ALagrangefunction}
	L_{\rho}\left(\boldsymbol{x},\boldsymbol{z},\boldsymbol{\lambda} \right):=& f\left(\boldsymbol{x}\right) + g\left(\boldsymbol{z}\right) + \boldsymbol{\lambda}^T \left( A\boldsymbol{x}+B\boldsymbol{z}-\boldsymbol{c} \right) \nonumber \\ 
	&  +\frac{\rho}{2}\left\|A\boldsymbol{x}+B\boldsymbol{z}-\boldsymbol{c} \right\|^{2},
	\end{align}
 	where $\rho > 0$ is a penalty parameter and
 	\begin{align*}
	 	\boldsymbol{\lambda}:=&\left[\vphantom{\left(\lambda_{p_{ij}}, \lambda_{q_{ij}}, \lambda_{w_{ij}},\lambda_{\theta_{ij}} \right)_{(i,j) \in \mathcal{L} \cup \mathcal{L}_{t}}}   \left(\lambda_{p,i}^{g}, \lambda_{q,i}^{g}  \right)_{(g,i)\in \mathcal{G}} ,  \left(\lambda_{p_{ij}}, \lambda_{q_{ij}}, \lambda_{w_{ij}}\right)_{(i,j) \in \mathcal{L} \cup \mathcal{L}_{t}} \right] \in \reals^{N_{\boldsymbol{\lambda}}},
 	\end{align*}
 	is the vector of dual variables associated with coupling constraints \eqref{eq:coupling}. The augmented Lagrangian in \eqref{eq:ALagrangefunction}
	is not separable in terms of sets of variables ($\mathcal{X}$ and $\mathcal{Z}$). Nonetheless, ADMM can be used to decouple these sets of variables, by using alternate minimizations over these sets. In particular, given the current iterates $\left(\boldsymbol{x}^{k},\boldsymbol{z}^{k},\boldsymbol{\lambda}^{k}\right)$, ADMM generates a new iterate $\left(\boldsymbol{x}^{k+1},\boldsymbol{z}^{k+1},\boldsymbol{\lambda}^{k+1}\right)$ as follows
	\begin{subequations}\label{eq:VanillaADMM}
		\begin{align}
			\boldsymbol{x}^{k+1} \in & \ \underset {\boldsymbol{x} \in \mathcal{X}} {\argmin } \ L_{\rho}\left(\boldsymbol{x},\boldsymbol{z}^{k},\boldsymbol{\lambda}^{k} \right), \label{eq:ADMMx} \\
			\boldsymbol{z}^{k+1} \in & \ \underset {\substack{\boldsymbol{z} \in \mathcal{Z}}} {\argmin } \ L_{\rho}\left(\boldsymbol{x}^{k+1},\boldsymbol{z},\boldsymbol{\lambda}^{k} \right), \label{eq:ADMMz} \\
			\boldsymbol{\lambda}^{k+1}=& \ \boldsymbol{\lambda}^{k}+\rho \left(A\boldsymbol{x}^{k+1}+B\boldsymbol{z}^{k+1}-\boldsymbol{c} \right). \label{eq:lupdate}
		\end{align}
	\end{subequations}
	More specifically, generators now solve
	\begin{subequations}\label{eq2:Genobj}
		\begin{align}
		D_{\rho,i}^{g}&\left(\boldsymbol{\lambda}^{g,k}_{i} \right)=\underset {\substack{\boldsymbol{x}_{gi}}} 
		{\mbox{minimize  }}  \sum_{g \in \mathcal{G}_{i}} \left( \vphantom{\left(p_{gi}-p_{i_{(i)}}^{g,k} \right)^{2}}  f_{gi} \left(p_{gi}\right) + \left\langle \boldsymbol{\lambda}^{g,k}_{i} , \boldsymbol{x}_{gi}\right\rangle \right. \nonumber \\
		& + \left. \frac{\rho}{2}\left( \left(p_{gi}-p_{i_{(i)}}^{g,k} \right)^{2} + \left( q_{gi} - q_{i_{(i)}}^{g,k}\right)^{2}  \right) \right)    \\
		& \hphantom{subject to} \text{subject to    } \text{\cref{eq0:Pminmax,eq0:Qminmax}},
		\end{align}
	\end{subequations}
	where $\boldsymbol{x}_{gi}:=[p_{gi},q_{gi}]$ and $\boldsymbol{\lambda}_{gi}:=\left[ \lambda_{p,gi},\lambda_{q,gi} \right] $, and branches (transmission lines, transformers) solve
	\begin{subequations}\label{eq:Lineobj}
		\begin{align}
		D^{\text{l}}_{\rho,ij} & \left(\boldsymbol{\lambda}^{k}_{ij},\boldsymbol{\lambda}^{k}_{ji}\right)=\underset {\substack{\boldsymbol{x}_{ij}^{\text{l}}}} 	{\mbox{minimize  }} \left\lbrace  \vphantom{\left( p_{ij} - p^{k}_{ij_{(i)}} \right)^{2}} \left\langle \left[\boldsymbol{\lambda}^{k}_{ij},\boldsymbol{\lambda}^{k}_{ji} \right], \boldsymbol{x}_{ij}^{\text{l}}\right\rangle + \right. \nonumber \\
		& \hspace{-0.6cm}  \left.  \sum_{(l,m) \in \left\lbrace (i,j) \cup (j,i) \right\rbrace } \frac{\rho}{2} \left( \vphantom{\left( p_{ij} - p^{k}_{ij_{(i)}} \right)^{2}}  \left( w^{k}_{l} - w_{l_{(lm)}} \right)^{2} + \right. \right. \nonumber \\
		 &  \left. \left.  \left( p_{lm} - p^{k}_{lm_{(l)}} \right)^{2} + \left( q_{lm} - q^{k}_{lm_{(l)}}\right)^{2}    \right) \right\rbrace   \\
		& \text{subject to \cref{eq1:Vminmax}, \cref{eq1:Pij,eq1:Qij,eq1:Pji,eq1:Qji,eq1:RSOC,eq1:anglediff}, \cref{eq0:linethermallimit}}, 
		\end{align}
	\end{subequations}
	where $\boldsymbol{x}_{ij}^{\text{l}}:=\left[ p_{ij},q_{ij},w_{i_{(ij)}},p_{ji},q_{ji},w_{j_{(ji)}}\right] $,  $\boldsymbol{\lambda}_{ij}:=\left[\lambda_{p_{ij}}, \lambda_{q_{ij}}, \lambda_{w_{ij}}\right]$ and $\boldsymbol{\lambda}_{ji}:=\left[\lambda_{p_{ji}}, \lambda_{q_{ji}}, \lambda_{w_{ji}}\right]$.
	On the other hand, buses solve
	\begin{subequations}\label{eq:Busobj}
		\begin{align}
		D^{\text{b}}_{\rho,i}&\left(\left(\boldsymbol{\lambda}^{g,k}_{i}\right)_{g \in \mathcal{G}_{i}}, \left( \boldsymbol{\lambda}^{k}_{ij}\right) _{j \in \mathcal{B}_{i}}\right)= \nonumber \\
		&\underset {\substack{\boldsymbol{z}_{i}}} {\mbox{minimize}} \left\lbrace \sum_{g \in \mathcal{G}_{i}} \left(\vphantom{\left(p_{gi}^{k+1}-p_{gi_{(i)}} \right)^{2}} - \left\langle \boldsymbol{\lambda}_{gi}^{k} , \left[ p_{gi_{(i)}},q_{gi_{(i)}}\right]  \right\rangle + \right. \right. \nonumber \\
		& \left. \left. \frac{\rho}{2}\left( \left(p_{gi}^{k+1}-p_{gi_{(i)}} \right)^{2} + \left( q_{gi}^{k+1} - q_{gi_{(i)}}\right)^{2}  \right) \right) +  \right. \nonumber  \\
		& \left. \hspace{-0cm}  \sum_{j\in \mathcal{B}_{i}}  \left(  - \left\langle  \boldsymbol{\lambda}_{ij}^{k} , \left[ p_{ij_{(i)}},q_{ij_{(i)}},w_{i}\right]  \right\rangle +  \right. \right.  \nonumber \\
		& \left. \left.  \frac{\rho}{2}\left( \left( p^{k+1}_{ij} - p_{ij_{(i)}} \right)^{2} + \left( q^{k+1}_{ij} - q_{ij_{(i)}} \right)^{2} \right) +    \right. \right.   \nonumber \\
		&   \left. \left. \left( \left( w_{i} - w^{k+1}_{i_{(ij)}} \right)^{2}  \right) \right) \vphantom{\sum_{g \in \mathcal{G}_{i}} \left( \left\langle \boldsymbol{\lambda}_{gi}^{k} , \left[ p_{gi_{(i)}},q_{gi_{(i)}}\right]  \right\rangle +\right)} \right\rbrace  , \\
		& \text{subject to    }  \text{\cref{eq1:KCL1,eq1:KCL2}}.
		\end{align}
	\end{subequations}
	where
	\begin{align*}
	\boldsymbol{z}_{i}:=\left[ \left( p_{gi_{(i)}},q_{gi_{(i)}} \right)_{(g,i)\in \mathcal{G}}, w_{i}, \left( p_{ij_{(i)}},q_{ij_{(i)}}\right)_{j\in \mathcal{B}_{i}}\right] .
	\end{align*}
	The advantages of applying the component-based decomposition to the alternative OPF formulation are that buses and generators admit closed-form solutions (see \cite{Peng_DOPFbalancedradial}).
	The primal residuals are defined as
	\begin{equation}
		\boldsymbol{r}^{k+1}=
		\left[r_{1}^{k+1}, \dots, r_{N_{\boldsymbol{\lambda}}}^{k+1} 	\right]  =A\boldsymbol{x}^{k+1}+B\boldsymbol{z}^{k+1}-\boldsymbol{c},
	\end{equation}
	and the dual residuals as
	\begin{equation}
	\boldsymbol{s}^{k+1}=
	\left[ s_{1}^{k+1}, \dots, s_{N_{\boldsymbol{\lambda}}}^{k+1}  \right]  =\rho A^{T} B\left( \boldsymbol{z}^{k+1}-\boldsymbol{z}^{k}\right).
	\end{equation}
	The algorithm in \eqref{eq:VanillaADMM} is terminated when
	\begin{align}
		\left\|\boldsymbol{r}^{k}\right\|  \leq \epsilon^{\text{pri}} \text{ and } \left\|\boldsymbol{s}^{k}\right\| \leq \epsilon^{\text{dual}},
	\end{align}
	where $\epsilon^{\text{pri}}$ and $\epsilon^{\text{dual}}$ are feasibility tolerances which are chosen using an absolute and relative criterion (see \cite{Boyd_ADMM}) as follows
	\begin{align}
		\epsilon^{\text{pri}}=&\sqrt{N_{\boldsymbol{\lambda}}}\epsilon^{\text{abs}}+\epsilon^{\text{rel}}\text{max}\left\lbrace \left\|A\boldsymbol{x}^{k}\right\|,\left\|B\boldsymbol{z}^{k}\right\|, \left\|\boldsymbol{c}\right\|  \right\rbrace, \\
		\epsilon^{\text{dual}}=&\sqrt{N_{\boldsymbol{x}}}\epsilon^{\text{abs}}+\epsilon^{\text{rel}}\left\|A^{T}\boldsymbol{\lambda}^{k}\right\|,
	\end{align}
	where $\epsilon^{\text{abs}} > 0$ and $\epsilon^{\text{rel}} > 0$ are absolute and relative tolerances respectively.
	The values of $\boldsymbol{r}^{k}$ and $\boldsymbol{s}^{k}$ indicate how distant the iterates are from a solution.\footnote{Note that the vanilla ADMM requires a central controller to check for convergence. However, if a central controller is unavailable, ADMM can in practice be run continuously over a fixed period of time, with no stopping criterion (see \cite{Kraning_ADMM}), which makes the scheme fully distributed.} If the sets $\mathcal{X}$ and $\mathcal{Z}$ are convex (which is the case for the SOCP-relaxed OPF) and problem \eqref{eq:Generalproblem} is feasible, ADMM is guaranteed to converge to an optimal point \cite{Boyd_ADMM}. The main objective of this paper is to establish accelerated variants for the algorithm in \eqref{eq:VanillaADMM} to ensure that these residuals decay quickly. To this end, this paper presents two accelerated methods in Section~\ref{sec:Acceleratedmethods} and an adaptive penalty parameter method in Section~\ref{sec:AdaptiveADMM}.	 	
	
	\section{Accelerated methods}\label{sec:Acceleratedmethods}
	
	Because of its simplicity, ADMM is a desirable way to solve \eqref{eq:Generalproblem}. However, since the OPF problem in high voltage transmission systems is inherently poorly conditioned (high inductance-to-resistance ratio), ADMM exhibits a poor performance, especially when high precision is required. In fact, ADMM is shown to have a convergence rate of $\mathcal{O}\left(\frac{1}{k} \right) $ \cite{He2012_ConvergenceofADMM}. This section presents two accelerated variants of ADMM, adapted to the SOCP-relaxed OPF setting.
		
	\subsection{Over-relaxed ADMM}\label{sec:OverrelaxedADMM}
		
	A popular variant of the algorithm in \eqref{eq:VanillaADMM} is the over-relaxed ADMM, which introduces a \emph{relaxation parameter} $\alpha \in \left(  0,2 \right)   $ and replaces each instance of $A\boldsymbol{x}^{k+1}$ in the $\boldsymbol{z}$ and $\boldsymbol{\lambda}$ updates in \eqref{eq:VanillaADMM} with
	\begin{align}
	\alpha A\boldsymbol{x}^{k+1} - \left(1-\alpha \right) \left( B \boldsymbol{z}^{k}-\boldsymbol{c}\right).
	\end{align}
	The over-relaxed ADMM is described in Algorithm~\ref{alg:Overrelaxed}. When $\alpha=1$, Algorithm~\ref{alg:Overrelaxed} and the vanilla ADMM in \eqref{eq:VanillaADMM} coincide. This method is analyzed in \cite{Eckstein1992_OntheDRSmethod} and \cite{Nishihara_GeneralanalysisofADMM}, and empirical studies in \cite{Eckstein1994_parallelADMM} show that over-relaxation with $\alpha \in \left[ 1.5,1.8\right]$ is more conducive to faster convergence.  	
	\begin{algorithm}[!t]
		\caption{Over-relaxed ADMM}
		\begin{algorithmic}[1]
			\scriptsize
			\STATE \parbox[t]{\dimexpr\linewidth-0.75cm}{\textbf{Initialization:} $k=1$, $\boldsymbol{\lambda}^{1} = \boldsymbol{0}$, $\boldsymbol{x}^{1} = \boldsymbol{0}$, $\boldsymbol{z}^{0} = \boldsymbol{0}$, $\rho > 0$, $\alpha \in \left(  1,2 \right]$, $\epsilon^{\text{abs}} = 10^{-6}$, $\epsilon^{\text{rel}} = 5 \times 10^{-5}$, and for all $i \in \mathcal{B}$, $\boldsymbol{z}_{i}^{\text{b},1}=\left[\left( 0.5\left( \underline{p}_{gi}+\overline{p}_{gi}\right) ,0.5\left( \underline{q}_{gi}+\overline{q}_{gi}\right)\right)_{(g,i)\in \mathcal{G}} ,1,\left( 0,0\right)_{j\in \mathcal{B}_{i}}\right]$\strut}			
			\WHILE {$\left\|\boldsymbol{r}^{k}\right\|  \geq \epsilon^{\text{pri}}$ \textbf{and} $\left\|\boldsymbol{s}^{k}\right\| \geq \epsilon^{\text{dual}}$} 
			\STATE \parbox[t]{\dimexpr\linewidth-0.75cm}{$\boldsymbol{x}^{k+1} \in  \ \underset {\boldsymbol{x} \in \mathcal{X}} {\argmin } \ L_{\rho}\left(\boldsymbol{x},\boldsymbol{z}^{k},\boldsymbol{\lambda}^{k} \right)$\strut}
			\STATE \parbox[t]{\dimexpr\linewidth-0.75cm}{$\hat{\boldsymbol{\lambda}}^{k+1}=\boldsymbol{\lambda}^{k}+\rho\left(\alpha-1 \right)\left(A\boldsymbol{x}^{k+1}+B\boldsymbol{z}^{k}-\boldsymbol{c} \right)  $\strut}
			\STATE \parbox[t]{\dimexpr\linewidth-0.75cm}{$\boldsymbol{z}^{k+1} \in  \ \underset {\boldsymbol{z} \in \mathcal{Z}} {\argmin } \ L_{\rho}\left(\boldsymbol{x}^{k+1},\boldsymbol{z},\hat{\boldsymbol{\lambda}}^{k+1} \right)$\strut}
			\STATE \parbox[t]{\dimexpr\linewidth-0.75cm}{$\boldsymbol{\lambda}^{k+1}=\hat{\boldsymbol{\lambda}}^{k+1}+\rho\left(A\boldsymbol{x}^{k+1}+B\boldsymbol{z}^{k+1}-\boldsymbol{c} \right)  $\strut}
			\STATE $k \leftarrow k + 1 $
			\ENDWHILE
		\end{algorithmic} 
		\label{alg:Overrelaxed}
	\end{algorithm}
	This method is also shown in \cite{Ghadimi2015_Optimalparameterselection} to improve convergence on quadratic problems.
	
	\subsection{Nesterov-type acceleration}\label{sec:Nesterov}
	
	In this section, the accelerated variant of ADMM is described in Algorithm~\ref{alg:Predictorcorrector}. The linchpin of this accelerated method is the predictor-corrector-type acceleration step containing an over-relaxation step,  initially proposed by Nesterov for gradient descent methods \cite{Nesterov2004_introductorylectures}. In this case the primal residual is unchanged but the dual residual is changed to
	\begin{align}
		\boldsymbol{s}^{k+1}=\rho A^{T} B\left( \boldsymbol{z}^{k+1}-\hat{\boldsymbol{z}}^{k}\right),
	\end{align}
	as in \cite{Goldstein2014_FastADMM}.
	\begin{algorithm}[!t]
		\caption{Predictor-corrector-type acceleration}
		\begin{algorithmic}[1]
			\scriptsize
			\STATE \parbox[t]{\dimexpr\linewidth-0.75cm}{\textbf{Initialization:} $k=1$, $\boldsymbol{\lambda}^{1} = \boldsymbol{0}$, $\boldsymbol{x}^{1} = \boldsymbol{0}$, $\boldsymbol{z}^{0} = \boldsymbol{0}$, $\rho > 0$, $\alpha^{1}=1$, $\eta \in \left( 0,1\right) $, $c^{1}=0$, $\epsilon^{\text{abs}} = 10^{-6}$, $\epsilon^{\text{rel}} = 5 \times 10^{-5}$, and for all $i \in \mathcal{B}$, $\boldsymbol{z}_{i}^{\text{b},1}=\left[\left( 0.5\left( \underline{p}_{gi}+\overline{p}_{gi}\right) ,0.5\left( \underline{q}_{gi}+\overline{q}_{gi}\right)\right)_{(g,i)\in \mathcal{G}} ,1,\left( 0,0\right)_{j\in \mathcal{B}_{i}}\right]$\strut}
			\WHILE {$\left\|\boldsymbol{r}^{k}\right\|  \geq \epsilon^{\text{pri}}$ \textbf{and} $\left\|\boldsymbol{s}^{k}\right\| \geq \epsilon^{\text{dual}}$} 
			\STATE \parbox[t]{\dimexpr\linewidth-0.75cm}{$\boldsymbol{x}^{k+1} \in  \ \underset {\boldsymbol{x} \in \mathcal{X}} {\argmin } \ L_{\rho}\left(\boldsymbol{x},\hat{\boldsymbol{z}}^{k},\hat{\boldsymbol{\lambda}}^{k} \right)$\strut}
			\STATE \parbox[t]{\dimexpr\linewidth-0.75cm}{$\boldsymbol{z}^{k+1} \in  \ \underset {\boldsymbol{z} \in \mathcal{Z}} {\argmin } \ L_{\rho}\left(\boldsymbol{x}^{k+1},\boldsymbol{z},\hat{\boldsymbol{\lambda}}^{k} \right)$\strut}
			\STATE \parbox[t]{\dimexpr\linewidth-0.75cm}{$\boldsymbol{\lambda}^{k+1}=\hat{\boldsymbol{\lambda}}^{k}+\rho\left(A\boldsymbol{x}^{k+1}+B\boldsymbol{z}^{k+1}-\boldsymbol{c} \right)  $\strut}
			\STATE \parbox[t]{\dimexpr\linewidth-0.75cm}{$c^{k+1}=\rho \left\|\boldsymbol{r}^{k+1} \right\|^{2}+\rho^{-1} \left\|\boldsymbol{s}^{k+1} \right\|^{2}   $\strut}
			\IF{$c^{k+1} < \eta c^{k}$}
			\STATE \parbox[t]{\dimexpr\linewidth-0.75cm}{$\alpha^{k+1}=\frac{\left(1+\sqrt{1+4\left( \alpha^{k}\right)^{2}} \right) }{2}$\strut}
			\STATE \parbox[t]{\dimexpr\linewidth-0.75cm}{$\hat{\boldsymbol{z}}^{k+1}=\boldsymbol{z}^{k+1}+\frac{\alpha^{k}-1}{\alpha^{k+1}}\left(\boldsymbol{z}^{k+1}-\boldsymbol{z}^{k} \right) $\strut}
			\STATE \parbox[t]{\dimexpr\linewidth-0.75cm}{$\hat{\boldsymbol{\lambda}}^{k+1}=\boldsymbol{\lambda}^{k+1}+\frac{\alpha^{k}-1}{\alpha^{k+1}}\left(\boldsymbol{\lambda}^{k+1}-\boldsymbol{\lambda}^{k} \right) $\strut}
			\ELSE
%			\STATE \parbox[t]{\dimexpr\linewidth-0.75cm}{$\alpha^{k+1}=1$\strut}
			\STATE \parbox[t]{\dimexpr\linewidth-0.75cm}{$\alpha^{k+1}=1$, $\hat{\boldsymbol{z}}^{k+1}=\boldsymbol{z}^{k+1}$, $\hat{\boldsymbol{\lambda}}^{k+1}=\boldsymbol{\lambda}^{k+1}$\strut}
			\STATE \parbox[t]{\dimexpr\linewidth-0.75cm}{ $c^{k+1} \leftarrow c^{k}$\strut}
			\ENDIF
			\STATE $k \leftarrow k + 1 $
			\ENDWHILE
		\end{algorithmic} 
		\label{alg:Predictorcorrector}
	\end{algorithm}
	This method was originally aimed at accelerating gradient descent-type (first-order) methods and was shown by Nesterov to achieve a complexity $\mathcal{O}\left(\frac{1}{k^{2}} \right) $, a rate which is proven to be optimal \cite{Nesterov2005_Smoothminimization,Nesterov2004_introductorylectures}. This method was first adapted to ADMM in \cite{Goldstein2014_FastADMM} and is further modified is this paper to suit the SOCP-relaxed OPF problem in which $f$ and $g$ are minimized over $x \in \mathcal{X}$ and $z \in \mathcal{Z}$ instead of over $x \in \reals^{N_{\boldsymbol{x}}}$ and $z \in \reals^{N_{\boldsymbol{z}}}$, respectively. More specifically, setting $\hat{\boldsymbol{z}}^{k+1}=\boldsymbol{z}^{k+1}$ and $\hat{\boldsymbol{\lambda}}^{k+1}=\boldsymbol{\lambda}^{k+1}$ in the \emph{restart} step on lines 12-13 of Algorithm~\ref{alg:Predictorcorrector} results in a faster convergence compared to $\hat{\boldsymbol{z}}^{k+1}=\boldsymbol{z}^{k}$ and $\hat{\boldsymbol{\lambda}}^{k+1}=\boldsymbol{\lambda}^{k}$.
	
	The convergence rate of $\mathcal{O}\left(\frac{1}{k^{2}} \right) $ can no longer be proven for Algorithm~\ref{alg:Predictorcorrector} because of the restart rule; nonetheless Algorithm~\ref{alg:Predictorcorrector} is still guaranteed to converge because problem \eqref{eq1:opf} is convex.\footnote{See \cite{Goldstein2014_FastADMM} for the proof.} The restart rule is needed in this OPF setting because the residuals do not decrease monotonically. Algorithm~\ref{alg:Predictorcorrector}'s acceleration potential is demonstrated empirically in Section~\ref{sec:Evaluation}. In all the numerical evaluations in Section~\ref{sec:Evaluation}, $\eta=0.999$ is used, as in \cite{Goldstein2014_FastADMM}. Note that in contrast to the over-relaxed ADMM method in the previous section, the predictor-corrector-type acceleration scheme is no longer fully distributed, as updating the combined residual $c^{k+1}$ on line 6 of Algorithm~\ref{alg:Predictorcorrector} requires the global computation of the primal and dual residuals.
		
%	\subsection{Bundle method}\label{sec:Bundlemethod}
		
	\section{Adaptive penalty parameter}\label{sec:AdaptiveADMM}
	
	The convergence of ADMM is in practice sensitive to the choice of $\rho$. A natural extension is to allow this parameter to vary at each iteration $k$. One such varying penalty parameter scheme, proposed in \cite{He2000_Selfadaptivepenalty} and \cite{WANG2001_Decompositiomethod}, updates $\rho$ based on the relative magnitudes of the primal and dual residuals as follows
	\begin{equation}\label{eq:RB}
		\rho^{k+1}:=
		\left\{\begin{aligned}
		& \rho^{k}\left(1+\tau^{\text{incr}} \right)  && \text{if } \left\|\boldsymbol{r}^{k+1}\right\| > \mu^{\text{incr}}\left\|\boldsymbol{s}^{k+1}\right\| , \\
		& \rho^{k}\left(1+\tau^{\text{dec}} \right)^{-1}  && \text{if } \left\|\boldsymbol{s}^{k+1}\right\| > \mu^{\text{decr}}\left\|\boldsymbol{r}^{k+1}\right\|, \\
		& \rho^{k} && \text{otherwise}, \\
		\end{aligned}
		\right.
	\end{equation}
	where $\tau^{\text{incr}} > 0$, $\tau^{\text{dec}} > 0$, $\mu^{\text{incr}} > 1$ and $\mu^{\text{decr}} > 1$ are parameters. The founding premise of this scheme is to balance the primal and dual residual magnitudes to within a factor of $\mu$ of one another as they both converge to zero. This varying penalty scheme is shown to improve the convergence of ADMM, in addition to making it less dependent on the initial choice of this parameter. However, this scheme is not suitable for a distributed setting as it relies on a central controller to compute the global residuals. Nonetheless, by introducing $\boldsymbol{\rho}^{k+1}:=\left[\rho_{1}^{k+1},\dots,\rho_{N_{\boldsymbol{\lambda}}}^{k+1} \right] $, where $\rho_{p}^{k+1}$ is a penalty parameter associated with each consensus constraint, the residual balancing scheme in \eqref{eq:RB} can be extended to the distributed setting as follows
	\begin{align}\label{eq:RBdynamic}
		\rho_{p}^{k+1}:=&
		\left\{\begin{aligned}
		& \rho_{p}^{k}\left(1+\tau^{\text{incr}} \right)  && \text{if } \left| {r}_{p}^{k+1}\right|  > \mu^{\text{incr}} \left| s_{p}^{k+1} \right|  , \\
		& \rho_{p}^{k+1}\left(1+\tau^{\text{dec}} \right)^{-1}  && \text{if } \left| s_{p}^{k+1}\right|  > \mu^{\text{decr}} \left| r_{p}^{k+1} \right| , \\
		& \rho_{p}^{k+1} && \text{otherwise}, \\
		\end{aligned}
		\right. \nonumber \\
		& \qquad \qquad \qquad \qquad \qquad \qquad p \in \left\lbrace 1,\dots, N_{\boldsymbol{\lambda}}\right\rbrace,
	\end{align}
	where ${r}_{p}^{k+1}$ and ${s}_{p}^{k+1}$ are the \emph{local} primal and dual residuals.\footnote{These local primal and dual residuals can be computed at the corresponding buses.} The generalization of \eqref{eq:RB} to distributed consensus problems has been explored in \cite{Song2016_Adaptivepenalty}. However, in contrast to the method in \cite{Song2016_Adaptivepenalty}, the update scheme \eqref{eq:RBdynamic} in this work is executed every $K_{f}=2$ iterations to avoid oscillations due to frequent changes in $\rho_{p}^{k+1}$. The end result, shown in Algorithm~\ref{alg:AdaptiveADMM} is an adaptive scheme that automatically tunes penalty parameters without central oversight, which is suitable for dynamic network topologies underlying distributed consensus problems. In practice, a system operator can conduct off-line vanilla ADMM simulations to get a ballpark figure of the penalty parameter values that lead to a fast convergence and these are then used as an initialization for Algorithm~\ref{alg:AdaptiveADMM}. In all the simulations in Section~\ref{sec:Evaluation} below, the parameters of Algorithm~\ref{alg:AdaptiveADMM} are set to $\tau^{\text{incr}} = 1$, $\tau^{\text{decr}} =0.5$, $\mu^{\text{incr}} = 10$ and $\mu^{\text{decr}} =100$. 
	\begin{algorithm}[!t]
		\caption{Adaptive consensus ADMM}
		\begin{algorithmic}[1]
			\scriptsize
			\STATE \parbox[t]{\dimexpr\linewidth-0.75cm}{\textbf{Initialization:} $\boldsymbol{\lambda}^{1} = \boldsymbol{0}$, $\boldsymbol{x}^{1} = \boldsymbol{0}$, $\boldsymbol{\rho}^{1} \succeq \boldsymbol{0}$, $\tau^{\text{incr}} > 0$, $\tau^{\text{dec}} > 0$, $\mu^{\text{incr}} > 1$, $\mu^{\text{decr}} > 1$, $\epsilon^{\text{abs}} = 10^{-6}$, $\epsilon^{\text{rel}} = 5 \times 10^{-5}$, and for all $i \in \mathcal{B}$, $\boldsymbol{z}_{i}^{\text{b},1}=\left[\left( 0.5\left( \underline{p}_{gi}+\overline{p}_{gi}\right) ,0.5\left( \underline{q}_{gi}+\overline{q}_{gi}\right)\right)_{(g,i)\in \mathcal{G}} ,1,\left( 0,0\right)_{j\in \mathcal{B}_{i}}\right]$, $K_{f}=2$, $k=1$ \strut}
			\WHILE {$\left\|\boldsymbol{r}^{k}\right\|  \geq \epsilon^{\text{pri}}$ \textbf{and} $\left\|\boldsymbol{s}^{k}\right\| \geq \epsilon^{\text{dual}}$} 
			\STATE \parbox[t]{\dimexpr\linewidth-0.75cm}{$\boldsymbol{x}^{k+1} \in  \ \underset {\boldsymbol{x} \in \mathcal{X}} {\argmin } \ L_{\boldsymbol{\rho}^{k}}\left(\boldsymbol{x},\boldsymbol{z}^{k},\boldsymbol{\lambda}^{k} \right)$\strut}
			\STATE \parbox[t]{\dimexpr\linewidth-0.75cm}{$\boldsymbol{z}^{k+1} \in  \ \underset {\boldsymbol{z} \in \mathcal{Z}} {\argmin } \ L_{\boldsymbol{\rho}^{k}}\left(\boldsymbol{x}^{k+1},\boldsymbol{z},\boldsymbol{\lambda}^{k} \right)$\strut}
			\STATE \parbox[t]{\dimexpr\linewidth-0.75cm}{$\boldsymbol{\lambda}^{k+1}=\boldsymbol{\lambda}^{k}+\boldsymbol{\rho}^{k}\left(A\boldsymbol{x}^{k+1}+B\boldsymbol{z}^{k+1}-\boldsymbol{c} \right)  $\strut}
			\IF{$\text{mod}\left(k,K_{f}=0 \right) $}
			\STATE \parbox[t]{\dimexpr\linewidth-0.75cm}{Locally update $\rho_{p}^{k+1}$ using \eqref{eq:RBdynamic} \strut}
			\ELSE
			\STATE \parbox[t]{\dimexpr\linewidth-0.75cm}{$\rho_{p}^{k+1} \leftarrow \rho_{p}^{k}$\strut}
			\ENDIF
			\STATE $k \leftarrow k + 1 $
			\ENDWHILE
		\end{algorithmic} 
		\label{alg:AdaptiveADMM}
	\end{algorithm}
			
	\section{Numerical evaluation}\label{sec:Evaluation}
	
	The two accelerated methods along with the adaptive ADMM scheme are evaluated on MATPOWER's case 5 \cite{MATPOWER} and on PEGASE test systems with up to 9241 buses \cite{Josz_ACdataMATPOWER}. The versatility and robustness of the methods are also assessed by evaluating them on NESTA's Case\_24\_ieee\_rts\_sad (small angle difference conditions) \cite{NESTA}, a test case which is particularly difficult to solve using the vanilla ADMM in \eqref{eq:VanillaADMM}, as shown in the column 2 of Table~\ref{tab:Alg_evaluation}.\footnote{The NESTA test cases are designed specifically to incorporate key network parameters such as line thermal limits and small angle differences, which are critical in optimization applications.}
%	The absolute and relative tolerances are set to $\epsilon^{\text{abs}} = 10^{-6}$ and $\epsilon^{\text{rel}} = 5 \times 10^{-5}$.
	The simulations are all conducted on a computing platform with 10 Intel Xeon E5-2687W v3 CPUs at 3.10GHz, 64-bit operating system, and 128GB RAM. In all simulations, OPTI \cite{OPTI} is used as a frontend modelling language along with IPOPT v3.12.5 \cite{IPOPT} as a backend solver for the convex branch subproblems in \eqref{eq:Lineobj}. Generator and bus subproblems are convex and admit closed-form solutions (see \cite{Peng_DOPFbalancedradial}). In Algorithms~\ref{alg:Overrelaxed} and~\ref{alg:Predictorcorrector}, $\rho=10$ is used for the active and reactive power consensus constraints \cref{eq:conspg,eq:consqg,eq:consp,eq:consq} and $\rho=100$ is used for the voltage consensus constraints \eqref{eq:consw}, \textit{i.e.} $\boldsymbol{\rho}:=\left[ \left( 10\right)_{2\left| \mathcal{G}\right|+2\left| \mathcal{L} \cup \mathcal{L}_{t} \right|}, \left(100 \right)_{\left| \mathcal{L} \cup \mathcal{L}_{t} \right| }  \right]$. This specific parameter tuning, which sets a higher value for the voltage consensus constraints, is shown in \cite{Mhanna2018_Componentbaseddecomposition} to significantly improve the convergence of vanilla ADMM and can also mean the difference between convergence and divergence in some cases. The performance of Algorithm~\ref{alg:Overrelaxed} for the three different values of $\alpha$ is listed in columns 2, 3 and 4 of Table~\ref{tab:Alg_evaluation}, whereas the performance of Algorithm~\ref{alg:Predictorcorrector} is listed in column 8.
%	\caption{Performance of the proposed methods measured in the number iterations to convergence (columns 2 to 9). Column 10 shows the percentage speed-up of the fastest method and the last column shows the maximum violation in the consensus constraints at the termination of the fastest method. Values in bold designate the fastest convergence.} 
	\begin{table*}[!t]
		%\normalsize
		\centering
		\caption{Performance of the proposed methods measured in the number iterations to convergence (columns 2 to 9). Column 10 shows the percentage speed-up of the fastest method and the last column shows $\text{max}\left( \boldsymbol{r}\right)$ at the termination of the fastest method.}
		%		\resizebox{\linewidth}{!}{%
		\begin{tabular}{| c | c | c | c | c | c | c | c | c | c | c |}
			\cline{2-9}
			\multicolumn{1}{c|}{} & \multicolumn{3}{c|}{Over-relaxed} & \multicolumn{3}{c|}{Over-relaxed \& Adaptive} &  \multicolumn{1}{c|}{Fast} & \multicolumn{1}{c|}{Fast 
				\& Adaptive} & \multicolumn{1}{c}{} & \multicolumn{1}{c}{} \\
			\cline{10-11}
			\multicolumn{1}{c|}{} & \multicolumn{3}{c|}{(Alg.~\ref{alg:Overrelaxed})} & \multicolumn{3}{c|}{(Alg.~\ref{alg:Overrelaxed} with \ref{alg:AdaptiveADMM})} &  \multicolumn{1}{c|}{(Alg.~\ref{alg:Predictorcorrector})} & \multicolumn{1}{c|}{(Alg.~\ref{alg:Predictorcorrector} with \ref{alg:AdaptiveADMM})} & \multicolumn{1}{c|}{Speed-up} & \multicolumn{1}{c|}{$\text{max}\left( \boldsymbol{r}\right) $}  \\ \cline{1-9}
			Test case   & $\alpha=1$  & $\alpha=1.5$  &  $\alpha=1.8$  & $\alpha=1$  & $\alpha=1.5$  &  $\alpha=1.8$ & $\eta=0.999$ & $\eta=0.999$ & (\%) &  \\\hline
			Case\_5	&	1,681	&	1,120	&	947	&	372	&	362	&	471	&	1,629	&	$\textbf{355}$	&	78.88	&	2.48E-04	\\\hline
			Case\_89\_PEGASE	&	2,677	&	1,857	&	1,600	&	1,220	&	900	&	$\textbf{877}$	&	1,911	&	915	&	67.24	&	1.59E-03	\\\hline
			Case\_1354\_PEGASE	&	723	&	645	&	696	&	494	&	586	&	972	&	818	&	$\textbf{467}$	&	35.41	&	9.70E-04	\\\hline
			Case\_2869\_PEGASE	&	906	&	679	&	857	&	567	&	679	&	961	&	854	&	$\textbf{560}$	&	38.19	&	2.28E-03	\\\hline
			Case\_9241\_PEGASE	&	2,737	&	3,700	&	10,000	&	820	&	10,000	&	10,000	&	2,868	&	$\textbf{737}$	&	73.07	&	7.78E-03	\\\hline
			Case\_24\_ieee\_rts\_sad	&	26,496	&	17,665	&	14,721	&	$\textbf{3,380}$	&	4,024	&	5,369	&	24,563	&	3,404	&	87.24	&	4.89E-04	\\\hline
			Case\_29\_ieee\_rts\_sad	&	1,569	&	1,259	&	1,134	&	$\textbf{314}$	&	730	&	718	&	1,233	&	329	&	79.99	&	2.00E-03	\\\hline			
		\end{tabular}
		\label{tab:Alg_evaluation}
	\end{table*}
	The adaptive consensus ADMM method (Algorithm~\ref{alg:AdaptiveADMM}) is overlaid with the over-relaxed ADMM (Algorithm~\ref{alg:Overrelaxed}) and the resulting scheme's performance is shown in columns 5, 6 and 7 of Table~\ref{tab:Alg_evaluation} for three different values of $\alpha$ and for $\boldsymbol{\rho}^{1}:=\left[ \left( 10\right)_{2\left| \mathcal{G}\right|+2\left| \mathcal{L} \cup \mathcal{L}_{t} \right|}, \left(100 \right)_{\left| \mathcal{L} \cup \mathcal{L}_{t} \right| }  \right]$.
%	\footnote{Note that for $\alpha=1$ this scheme coincides with the plain adaptive consensus ADMM in Algorithm~\ref{alg:AdaptiveADMM}.}
	The adaptive consensus ADMM method (Algorithm~\ref{alg:AdaptiveADMM}) is also overlaid with the predictor-corrector-type acceleration method (Algorithm~\ref{alg:Predictorcorrector}) and the resulting scheme's performance is shown in columns 9 of Table~\ref{tab:Alg_evaluation}, which is also initialized with $\boldsymbol{\rho}^{1}:=\left[ \left( 10\right)_{2\left| \mathcal{G}\right|+2\left| \mathcal{L} \cup \mathcal{L}_{t} \right|}, \left(100 \right)_{\left| \mathcal{L} \cup \mathcal{L}_{t} \right| }  \right]$. The second to last column in Table~\ref{tab:Alg_evaluation} shows the percentage speed-up of the fastest method relative to the vanilla ADMM (Algorithm~\ref{alg:Overrelaxed} with $\alpha=1$) and the last column shows the maximum violation in the consensus constraints \cref{eq:conspg,eq:consqg,eq:consp,eq:consq,eq:consw} at the termination of the fastest method.
	
	The main observations from Table~\ref{tab:Alg_evaluation} can be summarized as follows. The over-relaxed ADMM accelerates the convergence in all the cases for both $\alpha=1.5$ and $\alpha=1.8$, but without a clear-cut conclusion over which value of $\alpha$ is better in general. The predictor-corrector-type acceleration scheme also improves the convergence in most test cases. However, only when these accelerated subgradient schemes are overlaid with the adaptive ADMM method that substantial speed-ups are achieved. More specifically, the Fast \& Adaptive ADMM scheme generally exhibits the best performance but, in fact, the speed-up in the cases where the Fast \& Adaptive scheme is faster than the Over-relaxed \& Adaptive ADMM scheme is only marginal. The most notable acceleration is witnessed on NESTA's Case\_24\_ieee\_rts\_sad case on which the vanilla ADMM requires 26496 iterations to converge. The adaptive penalty parameter scheme reduces this iteration count to 3380, which is almost 8 times faster. This observation, along with the fact that the Over-relaxed \& Adaptive ADMM scheme is fully distributed, makes it the most attractive of the four algorithms even for the plain case ($\alpha=1$). This underscores the substantial contribution of the adaptive consensus ADMM to the overall speed-up. 
	
	% There can be other variants of the algorithms. One of them is to apply the adaptive method only when accelerated step in fast ADMM is successful.
	
	\section{Conclusion}\label{sec:Conclusion}
	
	This paper proposes and assesses two accelerated subgradient methods and an adaptive penalty parameter scheme for improving the convergence of ADMM on the component-based dual decomposition of the SOCP-relaxed OPF. Both accelerated subgradient methods are shown to improve convergence in most test cases but only when these accelerated subgradient schemes are overlaid with the adaptive ADMM that substantial speed-ups are achieved. This makes the adaptive ADMM scheme is a key contributor in substantially accelerating the convergence of ADMM. The end result, is a fast adaptive scheme that automatically tunes penalty parameters without central oversight, which is suitable for dynamic network topologies underlying distributed consensus problems.
	
	\bibliographystyle{IEEEtran}
	{\footnotesize
		\bibliography{AccADMM}}
	
\end{document}